\documentclass[
 reprint,
 preprintnumbers,
superscriptaddress,
 amsmath,amssymb,
 aps,
prb,
]{revtex4-2}
\newcommand{\thickhat}[1]{\mathbf{\hat{\text{$#1$}}}}

\usepackage{graphicx}
\usepackage{dcolumn}
\usepackage{bm}
\usepackage[hidelinks]{hyperref}
\usepackage[dvipsnames]{xcolor}
\usepackage{colortbl}
\usepackage{soul}
\usepackage{amsmath}
\usepackage[normalem]{ulem}
\hypersetup{
    colorlinks,
    linkcolor={red!50!black},
    citecolor={blue!50!black},
    urlcolor={blue!80!black}
}
\newcommand{\new}[1]{\textcolor{black}{{#1}}}
\newcommand{\cut}[1]{\textcolor{red}{{}}}
\newcommand{\newtwo}[1]{\textcolor{black}{{#1}}}
\newcommand{\cuttwo}[1]{\textcolor{red}{{}}}

\begin{document}

\preprint{LA-UR-24-29040}

\title{Group Conductivity and Nonadiabatic Born Effective Charges of Disordered Metals, Warm Dense Matter, and Hot Dense Plasma}

\author{Vidushi Sharma}
\email{vidushi@princeton.edu}
\affiliation{Theoretical Division, Los Alamos National Laboratory, Los Alamos, NM 87545, USA}
\affiliation{Center for Nonlinear Studies, Los Alamos National Laboratory, Los Alamos, NM 87545, USA}
\affiliation{Applied Materials and Sustainability Sciences, Princeton Plasma Physics Laboratory, Princeton, NJ 08540-6655, USA}
\author{Alexander J. White}
\email{alwhite@lanl.gov}
\affiliation{Theoretical Division, Los Alamos National Laboratory, Los Alamos, NM 87545, USA}

\noaffiliation

\date{\today}

\begin{abstract}
    The average ionization state is a critical parameter in plasma models for charged particle transport, equation of state, and optical response.  The dynamical or nonadiabatic Born effective charge (NBEC), calculated via first principles time-dependent density functional theory, provides exact ionic partitioning of bulk electron response for both metallic and insulating materials. The NBEC can be \cut{trivially} transformed into a ``group conductivity," \textit{i.e.}, the \textit{electron} conductivity ascribed to a subset of ions. We show that for disordered metallic systems, such as warm dense matter (WDM) and hot dense plasma, the static limit of the NBEC is different from the average ionization state\new{s}, but that the ionization state can be extracted from the group conductivity even in mixed systems. We demonstrate this approach using a set of archetypical examples, including cold and warm aluminium, low- and high- density WDM carbon, and a WDM carbon-beryllium-hydrogen mixture. 
\end{abstract}

\maketitle

Warm dense matter (WDM) physics is key to many complex systems \cite{Bonitz2020, Dornheim2023,Falk2018}. WDM is generated in the initial phases of an inertial confinement fusion experiment and forms the cores of planetary systems, ice giants and exoplanets \cite{Lorenzen2014,Prakapenka2021,Ehrenreich2020,Shawareb2024}. Its theoretical description poses a challenge.
Nearly every analytical model for materials properties in WDM or hot dense plasma (HDP) regimes involves the average ionization state, \textit{i.e.}, the effective or partial charge of ions, $Z_{\text{eff}}$ \cite{Murillo2013,Grabowski2020,Stanek2024,Haines2024}. This includes electronic \cut{conductivity} and thermal conductivity \cite{Cohen1950, Lee1984, Munro1994, Desjarlais2001, Perrot1987}, electron-ion relaxation rates \cite{Spitzer53}, X-ray scattering \cite{Nardi98,Lihua2012,Kraus2016,Döppner2023}, charged particle stopping power \cite{Ren2020, Malko2022,Li1993,Maynard1982,Gericke1996,Brown2005,Clauser2018,Casas2013}, inverse Bremsstrahlung absorption \cite{Stallcop1974,Turnbull2023,Turnbull2024}, as well as equation of state \cite{Slattery1980, Rogers2000, Clerouin2016}, ionic transport properties \cite{Ticknor2016,White2017}, and multi-species mixing rules \cite{Hansen1977, Bastea2005, Epperlein1986,Clerouin2020,Starrett2020,White2024}. 

In a weakly coupled plasma regime, where degeneracy and ion correlations are negligible, the Saha model can be applied \cite{Drake2018,Hu2016}. 
For highly degenerate plasmas, the Thomas-Fermi model can be utilized but it does not account for discrete electronic levels \cite{Drake2018,Ying1989}. When strong ion correlations emerge at liquid or solid densities, \textit{ab initio} multi-atom quantum mechanical simulations are imperative for determining $Z_{\text{eff}}$. This is challenging because $Z_{\text{eff}}$ is not a well-defined observable, \textit{i.e.}, there is no unique quantum operator which defines it.

There exists a gamut of charge partitioning schemes frequently employed in quantum chemistry, such as Mulliken \cite{Mulliken1955}, Bader \cite{Bader1985}, Hirshfeld \cite{Hirshfeld1977}, etc \cite{Wiberg2018}. However, in disordered metals and in the WDM regime, wherein a significant number of electrons behave as ``nearly-free", these assignments become substantially more difficult and arbitrary
\cite{Ertural2019}. An alternative approach is to extract $Z_{\text{eff}}$ from well-defined observables, \textit{e.g.}, the electrical conductivity \cite{White2022, Bethkenhagen2020} or the electronic density of states \cite{Hu2017, Driver2018}. However, these are \textit{bulk} properties of the system, not atomically resolved. This prevents the determination of $Z_{\text{eff}}$ distributions, or in the case of multi-component mixtures, the \cut{contribution from} \new{values for} different elements. In this letter, we show that by exploiting the nonadiabatic forces on atoms we can \new{efficiently} extract \cut{both an atomistically resolved conductivity $\sigma^G$ as well as an element-specific average charge, ${\bar Z}^{G}_{\text{eff}}$.}  \new{not only an element-specific average charge, ${\bar Z}^{G}_{\text{eff}}$, but also the fully atomistically resolved conductivity $\sigma^G$, and thus the macroscopic dielectric, reflectivity and absorption spectrum. This approach is generally applicable to models where the assignment of optical response to a subset of atoms is desired, \emph{e.g.}, surface atoms, defects, and moieties, molecular functional groups, and alloys.}

\begin{figure*}[t]
    \centering
    \includegraphics[width=1\linewidth]{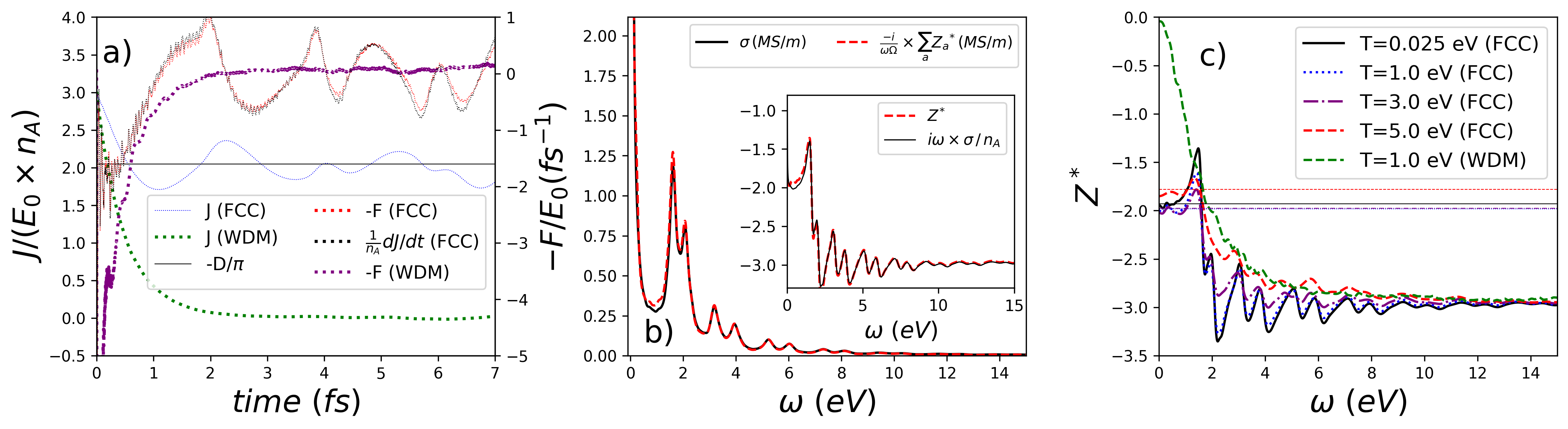}
  \caption{ Aluminium at solid-state density $\rho=2.7$ g/cm$^3$. 
  $a$) fcc at $k_BT = 0.025$ eV, disordered (WDM) at $k_BT = 1.0$ eV, on the vertical axes: (left) $J$ is scaled such that the signal is proportional to the number of electrons per atom, (right) negative of the force is scaled by $E_0$ yielding units of inverse time. Here, $-D/\pi$ is plotted for the fcc case.
  $b$) Current-- and force-- derived conductivity (Eq. \eqref{eq:groupcon}), and the \cut{current} \new{conductivity}-- and force -- derived average NBEC (inset, Eq. \eqref{eq:dynsumrule}) \new{for fcc Al at $k_BT = 0.025$ eV}.
  $c$) NBEC for fcc geometry at $k_B T = \{0.025, 1.0, 3.0, 5.0\}$ eV, and disordered (WDM) phase at $k_B T=1.0$ eV.}
    \label{fig:Al}
\end{figure*}

The static \emph{Born} (or transverse \cite{Ghosez1998}) effective charge tensor (BEC, ${\hat Z}^{*}_a$) is 
a well-defined quantum observable \cite{Born}. It is defined as the negative of the change in the atomic force vector (${\mathbf{F}}^{a}$) with respect to an applied electric field vector ($\mathbf{E}$), or equivalently, the change in the electronic polarization vector ($\mathbf{P}$) with respect to an atomic displacement vector ($\delta\mathbf{R}^{a}$) shifted by the bare nuclear charge ($Z_{a}$) :
\begin{align}
    {\hat Z}^{*}_a = -\dfrac{\partial \mathbf{F}_{a}}{\partial \mathbf{E}} = Z_a {\hat I} + \dfrac{\partial \mathbf{P}}{\partial \mathbf{R}^a}  ~,
\end{align}
where $\hat I$ is the identity matrix, and `$a$' indexes the atom. For gapped materials, \textit{i.e.}, insulators and semiconductors, ${\hat Z}^{*}$ can be calculated via perturbation theory \cite{Baroni2001}. However, for metallic systems and doped semiconductors, the electric polarization is ill-defined in the static limit \cite{Mahon2023}, and the static BEC has often been assumed to be ill-defined as well \cite{Resta2007}. Recently, Dreyer, Coh, and Stengel (DCS) used the concept of \emph{nonadiabatic} or \emph{dynamical} BEC (NBEC, ${\hat Z}^{*}_{a} (\omega + i\gamma) $), to extend BEC to metals \cite{Dreyer2022}.
Here $\omega$ is the frequency of the perturbing field and $\gamma$ is a small positive number 
. In the zero-frequency limit, $\omega \to 0$, the usual BEC is recovered \cite{Dreyer2022}. \cut{Wang \textit{et al.} provided a phenomenological framework for computing the finite-frequency NBEC for crystalline systems using Time-Dependent Density Functional Theory (TDDFT) \cite{Runge1984, Wang2022}. }

The DCS sum rule states that the sum of all NBEC's, divided by the cell volume $\Omega$, yields the ``Drude Weight",  
\begin{align}
\label{eq:DCS}
       \frac{1}{\Omega} \sum_a {\hat Z}^{*}_{a} ( 0 + i\gamma) \vert_ {\gamma\to 0} = - \dfrac{{\hat D}}{\pi} ~,
\end{align}
establishing a generalization of the acoustic sum rule applicable to both insulators (${\hat D}=0$) and metals (${\hat D} \ne 0$) \cite{Pick1970}. \textit{This} Drude weight is the ``truly-free" electron contribution to the electrical conductivity tensor (${\hat {\bf \sigma}}$). It is proportional to the number of electrons that, having ``high inertia" or weak-coupling to the ion lattice, do not respond to a rigid translation of the ionic sublattice or equivalently, do not relax to their ground-state after being kicked by an instantaneous uniform electric field \cite{Bellomia2020,Resta2018}. However, due to electron relaxation in disordered systems ${\hat D} \to 0$, and thus $\sum_a {\hat Z}_a^*(\omega)\vert_{\omega \to 0} \to 0$ as well, even in gap-less electronically-conducting systems such as disordered metals or WDM. The average of the diagonal of the NBEC tensor, ${\bar Z}^*(\omega \to 0)$, cannot \new{then} be directly taken as a measure of $Z_{\text{{eff}}}$ which must be finite for conductive systems. In these cases, electrons are only ``nearly-free" rather than ``truly-free" \cite{Bates1985}. A finite scattering time must be taken into account when determining $Z_{\text{{eff}}}$ from ${\bar Z}^*(\omega)$. As we will see, this can be done utilizing the \cut{low, but nonzero,}\new{finite} frequency information available from nonadiabatic dynamics. 

At finite frequency, the generalized DCS sum rule is written as,
\begin{align}
    \label{eq:dynsumrule}
      \frac{1}{N_a}\sum_a Z_a^* (\omega)  \equiv {\bar Z}^* (\omega) &=  i\frac{\omega}{n_a} \sigma(\omega) \equiv  \,  \, {\bar Z}_\text{CD}^* (\omega) ~, 
\end{align}
\cut{which can understood as an equivalence of the electronic response under changing reference frames \cite{Dreyer2022}.}
where $N_a$ is the number of atoms and $n_a$ is the atomic number density. We define ${\bar Z}_\text{CD}^* (\omega)$ as a conductivity-derived average (CDA-) NBEC. This is simply a consequence of the conservation of momentum \cite{Daligault2019, Marchese2024, Binci2021}; the time-derivative of the total canonical electron momentum must oppose the total change in the atomic forces minus the contribution due to the bare ion interacting with the electric field,
\begin{align}
    \label{eq:forcecurrentdef}
    \sum_a \delta{\bf F}^a(t) \vert_{{\bf R}^a(t)={\bf R}^a(0)} -Z_a {\bf E}(t)   = -\Omega \frac{d}{dt} {\bf J}(t) ~.
\end{align}
In this fixed-ion limit,  we assert that the momentum transferred to a particular atom or group of atoms (\textit{e.g.}, atoms of the same element) can be used to define a unique group ($G$) conductivity,    
    \begin{align}
    \label{eq:groupcon}
    \sum_{ a\in G} \frac{-i}{\Omega \omega}  Z_a^* (\omega)  &\equiv  {\bf \sigma}^G (\omega) ~,
    \\
      \sum_{G}  {\bf \sigma}^G (\omega) &\equiv \sigma^{FD}(\omega) = \sigma(\omega) ~.
\end{align}
Upon summing over all groups the force-derived conductivity ($\sigma^{FD}$) is equivalent to the conductivity from the generalized DCS sum rule in Eq. \eqref{eq:dynsumrule}. This transformation gives clear meaning to \new{the real and imaginary parts of} the NBEC at all frequencies, allowing for the extraction of atomistic details of the electron transport.

We calculate the NBEC and conductivity by simultaneously obtaining the time-dependent change in the atomic forces and current density in response to an instantaneous macroscopic external electric field pulse along the $x$-direction, $E_x(t) = \delta(t)E_{0,x}$. In this letter, we focus on isotropic systems; the \new{diagonals of the} NBEC and conductivity tensors are \cut{hence diagonal,} evaluated as,
\begin{align}
\label{eq:Zstarcalc}
    Z^{a\, *}_{xx} ( \omega) + Z_a   &= \int dt\, e^{{i\omega t -\frac{1}{4}\gamma^2t^2}} \Theta(t)  \dfrac{{F^{a}_x} (t) -{F^{a}_x} (0)}{E_{0,x}} ~,
    \\
        \sigma_{xx} ( \omega)  &= \int dt\, e^{{i\omega t -\frac{1}{4}\gamma^2t^2}} \Theta(t) \dfrac{{J_x} (t)}{E_{0,x}} ~.
\end{align}
While this approach to probe the linear-response conductivity has been previously reported \cite{Andrade2018, Kononov2022, White2022, Kononov2024}, the NBEC tensors have never been investigated in the WDM-HDP regime.

 \cut{The Time-Dependent Kohn-Sham (TD-KS-DFT)} Our approach for calculating $\mathbf{F}^{a} (t)$ and $\mathbf{J} (t)$ is outlined in Supplemental Material Sec. S1 \cite{Mermin1965,White2020, Sharma2023, White2022,Resta2018, Dreyer2022, Bellomia2020, Wang2022, Park1986,Goedecker1996,Hamann2013,Setten2018,Minary:2003,Monkhorst1976}.
  \new{Recently, Wang \textit{et al.} provided a phenomenological framework for computing the finite-frequency NBEC for crystalline systems, using Time-Dependent Density Functional Theory (TD-DFT), \cite{Runge1984} by calculating the current response to an instantaneous atomic displacement in crystalline systems \cite{Wang2022}.} \cut{To date only a single study has employed TD-KS-DFT to compute NBEC, relying on the current response to an instantaneous atomic displacement in crystalline systems \cite{Wang2022}.}
 Formally equivalent to Eq. \eqref{eq:Zstarcalc}, this approach lacks computational efficiency, especially for multi-atom (disordered) systems\new{.} \cut{ that do not benefit from an expansive Brillouin zone sampling, see Supplemental Material.}
 Our methodology allows for a simultaneous computation of $\sigma(\omega)$ and $Z^{*}_a (\omega)$ for all atoms 
 from a single TD-DFT simulation. \new{Notably, since we calculate only dynamical observables, any time-dependent electronic structure approach can be used.} Without loss of information,  from Eqs. \eqref{eq:dynsumrule} and \eqref{eq:groupcon}, we consider and compare only the real-parts of $\sigma (\omega)$,  $\sigma^{FD}(\omega)$, ${\bar Z}^* (\omega)$ and ${\bar Z}_\text{CD}^* (\omega)$, \textit{vide infra}.


Figure \ref{fig:Al}(a) shows the time-dependent current density ($\mathbf{J}$), and atomic forces ($\mathbf{F}$) in response to an instantaneous electric field pulse applied at $t=0$ for the cases of room-temperature fcc aluminium ($k_{B}T = 0.025$ eV), and warm dense disordered aluminium ($k_{B}T = 1.0$ eV).
A Drude weight of 2.04 (Eq. \eqref{eq:DCS}) is extracted for fcc Al in good agreement with previous results \cite{Wang2022}, see Supplemental Material. 
From Eq. \eqref{eq:forcecurrentdef}, $\frac{1}{n_a}\frac{\partial \mathbf{J}}{\partial t}$ is shown to be directly proportional to the average force; interestingly, unlike the fcc case $\mathbf{J}(t)$ in WDM decays to zero due to disorder.

In Fig. \ref{fig:Al} (b) the conductivity calculated from the current density ($\sigma (\omega)$) is compared with the conductivity derived from the \cut{Drude weight}\new{NBEC} ($\sigma^G (\omega)$, Eq. \eqref{eq:groupcon} used in conjunction with Eq. \eqref{eq:DCS}) for fcc Al.
The Drude contribution is added as a Lorentzian centered at $\omega = 0$ with a tenth of the artificial broadening (\textit{vide infra}, Supplemental Material S2). \new{The true width of this peak depends on lattice vibrations and their coupling to the electrons, which we do not consider here.}
Similarly, the inset of Fig. \ref{fig:Al}(b) shows an agreement between ${\bar Z}^* (\omega)$ and ${\bar Z}_\text{CD}^* (\omega)$, Eq. \eqref{eq:dynsumrule}.
Finally, Fig. \ref{fig:Al} (c) displays ${\bar Z}^* (\omega)$ computed for a range of electronic temperatures in the fcc Al structure along with disordered phase Al at $k_B T=$ 1.0 eV. The electron transport properties in Al are largely unaffected by the increased electronic temperatures; the Drude weight remains $\sim 2$ for up to $k_BT=5$ eV. 
Going from perfectly crystalline (blue-dotted) to disordered (green-dashed) eliminates the Drude weight contribution and $\bar{Z}^* \to 0$ in the static limit. 
This illustrates that $Z_{\text{eff}}$, which should be finite, and $\bar{Z}^* (\omega \to 0)$ are inherently different quantities in \cut{WDM} \new{disordered} systems. 
\begin{figure}[h]
    \centering
    \includegraphics[width=0.9\linewidth]{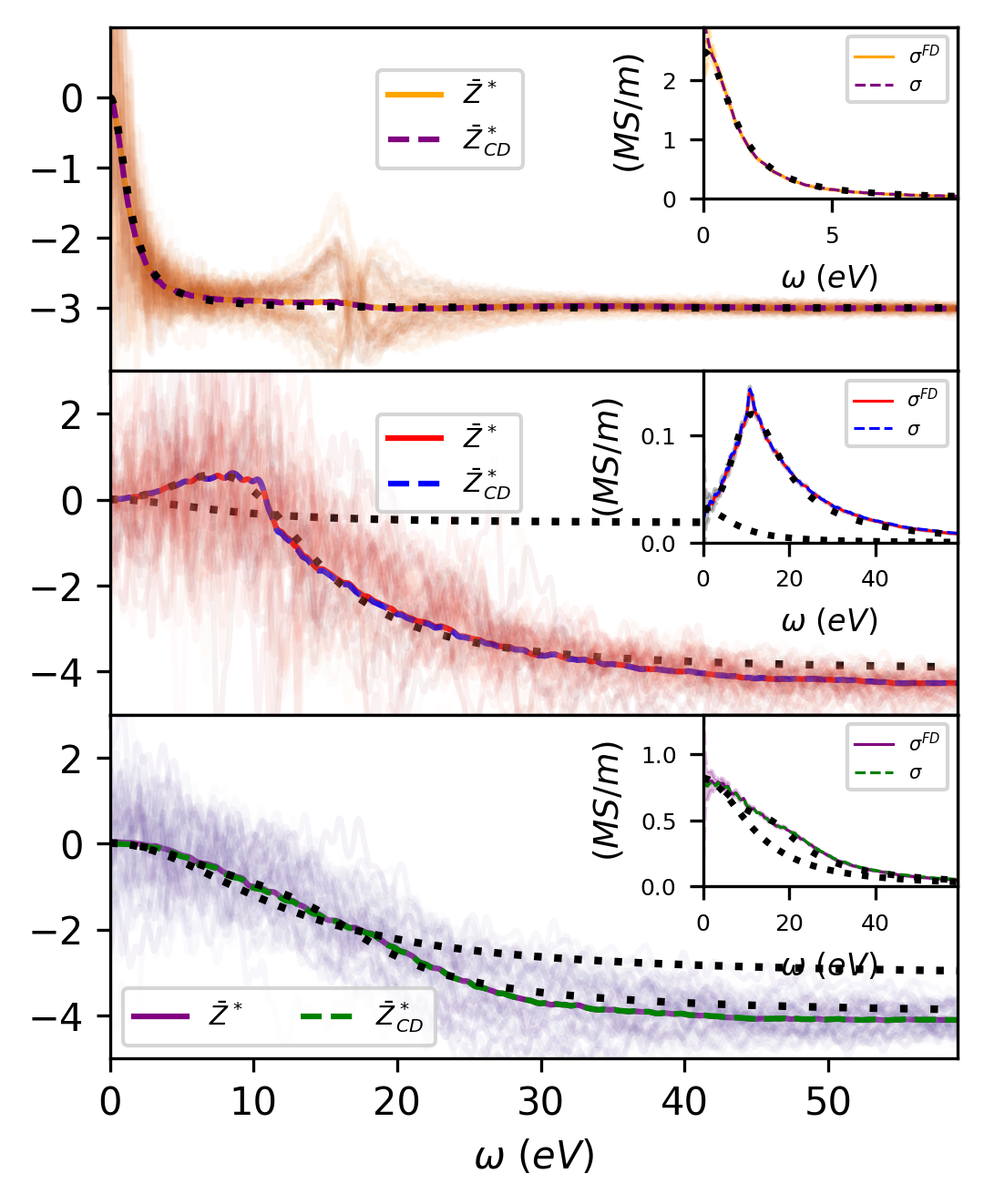}
    \caption{The top panel shows the NBEC (inset- conductivity) for WDM disordered Al at $k_B T = 1.0$ eV, while the middle and lower panels show  low-density ($0.5$ g/cm$^3$) and solid-density ($3.5$ g/cm$^3$) warm dense carbon systems at $k_B T=1.0$ eV.  \new{The gradient-shaded lines show individual contribution of atoms to the spectra. (Dense/Sparse) dotted black lines are the (Drude-only/full) fits. Insets show $\pm 1$ standard error of the force-derived conductivity as shaded bands.}
     }
    \label{fig:C}
\end{figure}

\begin{figure*}[t]
\
    \centering
    \includegraphics[width=1.0\linewidth]{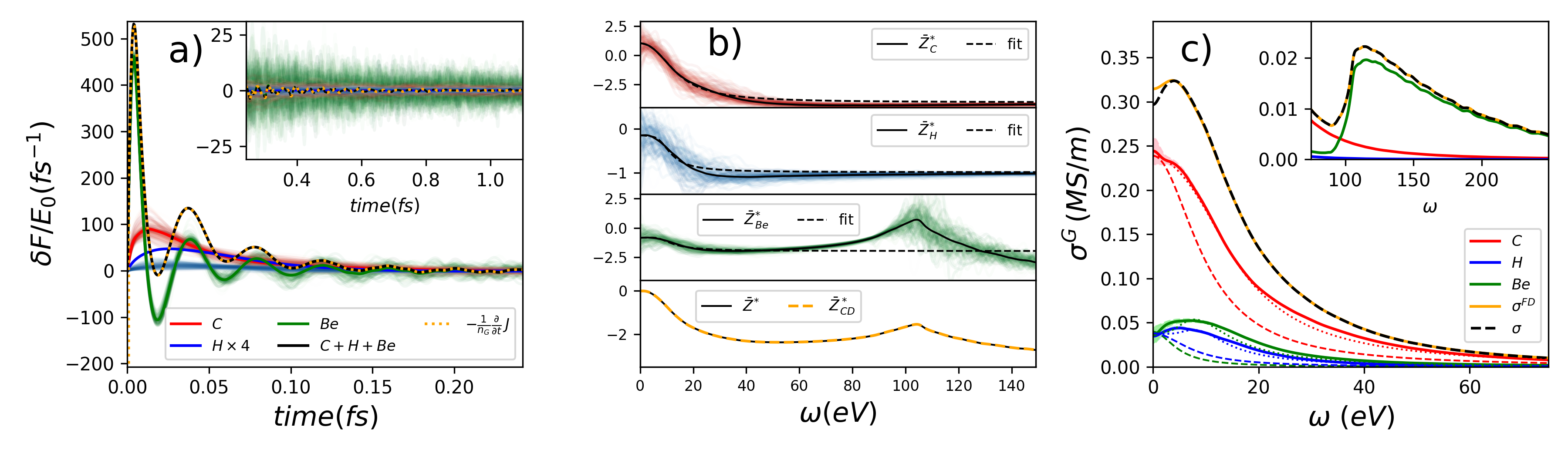}
    \caption{Carbon Hydrogen Beryllium mixture, 1.37 g/cm$^3$, $k_B T =5 $ eV. $a$) The time-dependent individual and element average forces, $F_x(t)-F_x(0)$, sum of the averages (black), and $3\times$ scaled time-derivative of the current-density per atom (orange). $b$) NBEC for different elements with average and \cut{Drude-Smith} \new{modified Drude-Lorentz} fit, bottom-average of all NBEC (black) and CDA-NBEC (orange). $c$) The total FD conductivity, along with the components from each element. The inset of $a$ (resp. $c$) extends the time (resp. frequency) domain. }
    \label{fig:CHBE}
\end{figure*}

We now propose the extraction of the average ${Z}^G_{\text{eff}}$ from ${\bar{Z}}^{*}_G (\omega)$ in warm dense matter. As was used for the group conductivity, $G$ indexes a given group of ions. We follow a typical approach of fitting the conductivity to a Drude\cut{-Smith} model \cut{\cite{Cocker2017}}. However, we also need to account for static charge transfer between groups, $\Delta Z^G_{CT}\equiv{\bar Z}_G^* (0)$\new{, and bound-free transitions}. For group conductivity, the f-sum rule is now modified to include both the average bare-ion charge of the group, $Z_G$, and $\Delta Z^G_{CT}$,
\begin{align}
\label{sum-rule}
\frac{2}{\pi}\int^\infty_0 d\omega \,  {\bf \sigma}^G (\omega) =  \big(Z_G + \Delta Z^G_{CT}\big) n_G ~,
\end{align}
 where $n_G$ is the number density of ions in group $G$. This follows from a straightforward application of the Kramers-Kronig relations to the force response. \new{We use a modified Drude-Lorentz model, Supplemental Material Sec. S4, to fit the ${\bf \sigma}^G (\omega)$ and extract $Z_G$.} \cut{We thus modify the Drude-Smith model to account for charge transfer:}
\cut{Here $\tau_G$ is the scattering time for the group, $c_G$ is a parameter that suppresses the low-frequency contribution due to back-scattering \cite{Cocker2017}. }
If $G$ includes all atoms of the system then $\Delta Z^G_{CT} = 0$. Drude-like behavior is inherently classical and emerges from quantum systems only when considering many ions. We identify that this approach is not applicable when $G$ includes too few atoms. Therefore working in the large system size limit for mixed systems, we fit the conductivity for $G$ comprising all atoms of a given elemental species. 

In Fig. \ref{fig:C} we show the average NBEC spectrum, ${\bar Z}^* (\omega)$ (solid and dashed lines), along with individual atom contributions (gradient-shaded), and (insets) $\sigma^G(\omega)$ for WDM systems. 
These average force-derived quantities are equivalent to their counterparts calculated from the current density, see Supplemental Material S3. While ${\bar Z}^* (\omega)$ for Al is an essentially featureless-dissipative spectrum; the individual $Z^{*}(\omega)$ show transition peaks between $\hbar\omega= 10$ to $20$ eV which average to zero\new{, indicating no net momentum transfer from the electrons}. \new{These occur near the forbidden Al$^{2+}$ $3s$ to $3d$ or $4d$ transitions \cite{Kelleher08}.}  \cut{These are the valence -- continuum transitions, as only the $3s^23p^1$ electrons are non-frozen (as part of the defined pseudopotential), with a very shallow Cooper minimum \cite{Cooper1962}.} \cut{More prominent non-Drude transitions have been seen for $\hbar\omega > 20$ eV from simulations including explicit $2s^22p^6$ electrons \cite{Witte2017}.} 
Here, $\sigma(\omega)$ or $ \sigma^{FD}(\omega)$ \cut{is fit} \new{fit readily} to a Drude model yielding \cut{$Z_{\text{eff}}= 2.9$} \new{$Z_{\text{eff}}= 3.0$}  \cut{$\tau = 0.47$} \new{with a scattering time, $\tau = 0.48$} fs. 
To make contact with the plasma physics models, we compare with the average-atom DFT code \texttt{Tartarus} \cite{Starrett2017} which furnishes two definitions of $Z_{\text{eff}}$, $Z^{AA,1}_{\text{eff}} =2.0$ from the population of KS states with non-negative energy, and $Z^{AA,2}_{\text{eff}}=3.0$ from the population of KS states similar to the free-electron states \cite{Starrett2019}.

The low-density (0.5 g/cm$^3$) carbon exhibits clear ``bound-free" transitions, making fitting the NBEC spectrum to a Drude model a difficult task.\cut{ The NBEC spectrum is composed of overlapping ``dispersive" line-shapes and the high frequency asymptotic limit approaches the bare (minus frozen) ion charge of $\sim -4$. By transforming to the group conductivity (Eq. \ref{eq:groupcon}, insets), the peaks become ``absorptive'' and separate.} \cut{Thus we} \new{We} fit \cut{the low-frequency portion of} the group conductivity to the \new{modified} Drude\new{-Lorentz} model and extract a \cut{$Z_{\text{eff}} = 0.58$ and $\tau = 0.22$ fs} \new{$Z_{\text{eff}} = 0.55$}; compared to \texttt{Tartarus} which produces $Z^{AA,1}_{\text{eff}} =0.59$ and $Z^{AA,2}_{\text{eff}}=0.74$. \new{The remainder of the spectral weight is in a Lorentz peak at $\omega_1 = 11.6$ eV, proportional to the number of bound electrons.} 
On the other hand, solid-density carbon has a nearly featureless Drude-like conductivity and NBEC spectrum. Looking at ${\bar Z}^* (\omega)$, one may expect that the system characteristics follow a Drude behavior over a large frequency range, yielding $Z_{\text{eff}} \sim 4$. However, a fit to $\sigma^G (\omega) $ reveals a distinct \cut{shoulder starting at $\hbar\omega \sim 10$ eV.} \new{Lorentz peak centered at $\omega_1 = 16.5$ eV. This leads to a $Z_{\text{eff}} = 3.1$, while} 
\cut{Therefore constraining the fit to low frequencies, we obtain a $Z_{\text{eff}} = 2.8$ and $\tau = 0.075$ fs;} \texttt{Tartarus} produces \cut{$Z^{AA,1}_{\text{eff}}=3.45$ and $Z^{AA,2}_{\text{eff}}=1.47$} \new{$Z^{AA,1}_{\text{eff}}=1.47$ and $Z^{AA,2}_{\text{eff}}=3.45$}.
\cut{We must emphasize at this point that our formalism of nonadiabatic electron dynamics gives us access to scattering (relaxation) times which cannot be generated using static perturbation theory-- based calculation.}
The distribution of $Z^*(\omega)$ is significantly larger for the high-density carbon compared to the low-density, as evidenced by the purple and red gradient-shaded lines respectively, in Fig. \ref{fig:C}.
\new{In all cases the modified Drude-Lorentz model, and the Drude only contribution proportional to $Z_{\text{eff}}$, based on fitting the group conductivity's, are plotted alongside the calculated results for both ${\bar Z}^* (\omega)$ and $\sigma^{FD} (\omega) $.}

\new{Thus-far we have only considered the group conductivity containing all atoms, which is identical to the regular conductivity. Now we consider the novel case where we can extract more detailed information by averaging only over subsets of atoms. Specifically we will examine the group conductivity containing atoms of different element types.}
\cut{As a final example, we} \new{We} present a single warm dense mixture of equal parts hydrogen, beryllium and carbon, elements integral to inertial confinement fusion ablators and recent experiments \cite{Hurricane2023,Jiang2023}, at a density of 1.37 g/cm$^3$ and $k_B T = 5$ eV.
This system comprises a large simulation box with 128 atoms of each element and high temperature at a relatively low density; physical conditions that challenge the deterministic Kohn-Sham approach \cite{White2020, Sharma2023}.
Hence we employ a recently developed mixed stochastic-deterministic TD-DFT to converge the simulation \cite{White2020, White2022}.
The force response of each ion (inset) and their element-wise and all-ion averages are shown in Fig. \ref{fig:CHBE}(a). 
As expected, the average force for all ions nearly exactly opposes the time-derivative of the current density per atom (from Eq. \eqref{eq:forcecurrentdef}). Due to the explicit treatment of Be$-1s$ electrons in the pseudopotential, the Be response features long-lived oscillations which quickly average to zero, but individually persist.
In Fig. \ref{fig:CHBE}(b), the  NBEC's for ions of each species are shown along with ${\bar Z}(\omega)$ and ${\bar Z}^{CD}(\omega)$. Figure \ref{fig:CHBE} (c) shows the group conductivity for each element type, the sum of the contributions, $\sigma^{FD}$, and the \new{regular} conductivity $\sigma$ \cut{are shown}. 
The peak in the conductivity starting at $\hbar\omega \sim 100$ eV is solely attributed to Be \new{$1s \to$ continuum} transitions, though it overlaps with the tail of the carbon contribution. A \cut{Drude-Smith} \new{modified Drude-Lorentz} fit\cut{s to low frequency segments} of each $\sigma^G(\omega)$ yields \cut{${\bar Z}_{\text{eff}}^{C}= 0.94$, ${\bar Z}_{\text{eff}}^{H}=1.00$, and  ${\bar Z}_{\text{eff}}^{Be}=2.23 $ along with $\tau_C=0.12$ fs, $\tau_H=0.16$ fs, and $\tau_{Be}=0.17$ fs, and $c_C=0.0$, $c_H=-0.91$, and $c_{Be}=-0.99$.} \new{${\bar Z}_{\text{eff}}^{C/H/Be}=2.36 / 0.57 /1.13.$} The NBEC calculations indicate charge transfer occurring between the groups, with \cut{$\Delta Z^{C}_{CT}=+1.11$ , $\Delta Z^H_{CT}=-0.18$, and  $\Delta Z^{Be}_{CT}=-0.94$} \new{$\Delta Z^{C/H/Be}_{CT}=+1.05/-0.16/-0.83$ electrons}, which follows from the electronegativity of each species. \new{Using a  chemical potential matching procedure \cite{Starrett2020}, \texttt{Tartarus} produces $Z^{AA,1,C/H/Be}_{\text{eff}}=2.78/1.0/2.0$ and $Z^{AA,2,C/H/Be}_{\text{eff}}=1.05/0.45/1.26$.}

We present a new formalism for a more efficient calculation of the $Z_a^{*}(\omega)$ for all atoms in a unit cell from TDDFT simulations.
\textit{Ab initio} calculations of $Z_a^{*}(\omega)$ could help develop a more complete understanding of the initial nonequilibrium ion-velocity distribution resulting from pulsed-laser excitations in the extreme ultraviolet and soft X-ray regimes. Transforming $Z_a^{*}(\omega)$ to the force-derived \new{conductivity,} $\sigma^{FD}(\omega)$\new{,} provides a logical interpretation of the high frequency and imaginary parts of $Z_a^{*}(\omega)$ \new{and the potential to partition the optical response into atomic groups, \textit{e.g.}, by element types}. Our simulations also provide the first numerical validation of the DCS sum rule beyond the static limit.
In the warm dense matter regime described here, averaging $Z_a^{*}(\omega)$ over a sufficiently large number of atoms allows determination of ${ Z}^G_{\text{eff}}$ for a subgroup of (or all)  atoms. These results tend to fall between the disparate definitions from an average atom code \cite{Starrett2020}. When applied to the mixed C/H/Be system, we see a charge transfer between the element groups, which unveils a new complexity to the electronic structure of warm dense matter mixtures. This could play a significant role in the mixing of equation-of-state or conductivity tables and model development. Moreover, during the generation of mixed conductivity spectra, the weighted contributions of different species could be compared against the group conductivities, $\sigma^G(\omega)$, generated by our proposed approach. 

\newtwo{The calculations in this letter are based on TD-DFT using approximate adiabatic exchange-correlation functionals \cite{Lacombe2023}. More research into the dynamical exchange-correlation effects, and development of accelerated nonequilibrium time-domain methods \cite{Bonitz24, Dornheim23}, and their effect on NBEC is warranted. Since the NBEC and group conductivities are well-defined observables, they can in principle be calculated using any excited-state electronic structure approach. However, the extraction of $Z^\text{G}_\text{eff}$ from the NBEC is based on a further assumption, \textit{i.e.,} the Drude-Lorentz picture, and is therefore not a unique estimation of the charge. We have shown that, at least, the NBEC model consistently relates $Z^\text{G}_\text{eff}$ to both the average ionic forces and the nearly-free electron conduction.}

\begin{acknowledgments}
This work was supported by the U.S. Department of Energy through the Los Alamos National Laboratory (LANL). Research presented in this article was supported by the Laboratory Directed Research and Development program, projects number 20230322ER and 20230323ER, and the Institute for Material Science, projects number 20248109CT-IMS, of LANL. We acknowledge the support of the Center for Nonlinear Studies (CNLS). This research used computing resources provided by the LANL Institutional Computing and Advanced Scientific Computing programs. Los Alamos National Laboratory is operated by Triad National Security, LLC, for the National Nuclear Security Administration of U.S. Department of Energy (Contract No. 89233218CNA000001).
\end{acknowledgments}

\bibliography{main.bib}



\clearpage
\clearpage 
\setcounter{page}{1}
\renewcommand{\thetable}{S\arabic{table}}  
\setcounter{table}{0}
\renewcommand{\thefigure}{S\arabic{figure}}
\setcounter{figure}{0}
\renewcommand{\thesection}{S\arabic{section}}
\setcounter{section}{0}
\renewcommand{\theequation}{S\arabic{equation}}
\setcounter{equation}{0}
\onecolumngrid

\begin{center}
\textbf{Supplemental Material for\\\vspace{0.5 cm}
\large Group Conductivity and Dynamical Born Effective Charges of Disordered Metals, Warm Dense Matter and Dense Plasma Mixtures\\\vspace{0.3 cm}}

Vidushi Sharma$^{1,2,3}$ and Alexander J. White$^{1}$

\small

$^1$\textit{Theoretical Division, Los Alamos National Laboratory, Los Alamos, NM 87545, USA}

$^2$\textit{Center for Nonlinear Studies (CNLS), Los Alamos National Laboratory, Los Alamos, NM 87545, USA}

$^3$\textit{Applied Materials and Sustainability Sciences, Princeton Plasma Physics Laboratory, Princeton, NJ 08540-6655, USA}

(Dated: \today)
\end{center}

\section{\label{sec:AppendixA} Time-Dependent Kohn-Sham approach to linear response conductivity and nonadiabatic Born effective charge}
Since conductivity and NBEC are quantum mechanical observables with well-defined operators, they can be calculated using any time-dependent or excited-state electronic structure method. Here we calculate and analyze the NBEC and group conductivity for bulk systems using time-dependent Kohn-Sham density functional theory \new{(TD-DFT)}. \new{TD-DFT is the current state of the art excited-state method for large systems with hundreds to thousands of atoms and for warm dense matter, where high temperatures lead to an additional increase in computational cost. Calculations including dynamic correlation, such as, GW and time-dependent Bethe-Salpeter, may improve the results but are beyond the scope of this letter.} We utilize atomic units such that the mass of electron, $m_e$, and the reduced Planck's constant $\hbar$ are unity. Within linear response, the time-dependent NBEC tensor is defined as the response function of the negative of the atomic force vector with respect to the electric field vector,  
    \begin{align}
    d {\bf F }^a (t,{\bf q})  \equiv  - \int\int dt' d{\bf q }' \, {\hat { Z}^{*}_a(t,t',{\bf q},{\bf q'}} )\cdot d{\bf E}({\bf q}',t') ~,
    \end{align}
while the conductivity tensor is  defined by a similar equation with the negative of the force replaced by the current density,  
    \begin{align}
    d {\bf J} (t,{\bf q})  \equiv \int\int dt' d{\bf q }' \, {\hat {\bf \sigma}(t,t',{\bf q},{\bf q'}} )\cdot d{\bf E}({\bf q}',t') ~.
    \end{align}
Here ${\bf q}$ is the perturbation wavevector. If we consider a macroscopic instantaneous electric field pulse along an arbitrarily chosen $x-$direction, $d E_x(q',t')=E_0 \delta(q')\delta(t')$, we can obtain the macroscopic NBEC  / conductivity via 
\begin{align}
\label{Zstarcalc}
    {\hat Z}^{*}_{a,y,x} (\omega) &\equiv  \mathcal{F}\big[\Theta(t)\,d{F}^{a,y} (t)/E_0 \big](\omega) ~,
    \\
        {\sigma}^{*}_{y,x} (\omega) &\equiv  \mathcal{F}\big[\Theta(t)\,d{J^y} (t)/E_0 \big](\omega) ~,
    \end{align}
where $\mathcal{F}[...]$ denotes the Fourier transform from the forward time $t$ to angular-frequency $\omega$; the Heaviside function, $\Theta(t)$ is included to ensure causality. 

To calculate the time-dependent forces and current from `adiabatic' TD-DFT \cut{(TD-KS-DFT)} we solve the equation of motion for the KS Bloch orbitals:
\begin{align}
    \frac{i\partial}{\partial t} u^{\bf k}_b({\bf r}, t) = {\thickhat{H}}^{\bf k}_{KS} (t) u^{\bf k}_b( t) ~,
    \end{align}
\begin{align}
    {\thickhat{H}}^{\bf k}_{\text{KS}} (t) = -\dfrac{1}{2} ({\thickhat \nabla} + i {\bf A}(t) + i {\bf k})^2 + V_{\text{XC},H} ({\bf r}, \rho(t)) +  V_{ei, \text{LPP}} ({\bf r}, {\bf R}) + {\thickhat{V}}^{\bf k}_{ei,\text{NLPP}} ( {\bf R}) ~.
\end{align}
Here the term `adiabatic' refers to the dependence of the exchange-correlation (plus Hartree) potential, $\hat V_{\text{XC},H}$, on only the instantaneous time-dependent density, $\rho(t)$; ${\bf r}$ is the real-space electron position vector while ${\bf R}$ is the ion position vector, ${\bf k}$ is the k-point associated with this Bloch orbital. When using pseudopotentials (PP), the electron-ion interaction has a local part, $V_{ei, \text{LPP}} ({\bf r}, {\bf R})$, and possibly a nonlocal part, ${\thickhat{V}}^{\bf k}_{ei,\text{NLPP}} ( {\bf R})$. The time-dependent density (density matrix) is given by:
\begin{align}
\label{DM}
\rho({\bf r},t) &= \delta({\bf r},{\bf r'} )  \rho({\bf r},{\bf r'}t) = \delta({\bf r},{\bf r'} )\sum_{{\bf k}, b} f({\varepsilon^{\bf k}_b, \mu, T}) u^{{\bf k}}_b ({\bf r},t) u^{*,{\bf k}}_b ({\bf r'},t) ~, \quad \text{where}
\\
f({\varepsilon, \mu, T}) &:= \frac{1}{1 + e^{(\varepsilon-\mu)/k_{B}T}} ~.
\end{align}
In Mermin's extension of TD-\cut{KS-}DFT, the finite electron temperature \cite{Mermin1965} $T$, enters only through the Fermi-Dirac occupations, which depend on the equilibrium Mermin-DFT eigenenergies, $\varepsilon$, and the electron chemical potential, $\mu$. For mixed stochastic-deterministic KS-DFT, the density matrix contains also the complementary stochastic vectors, which take a similar low-rank form to Eq. \eqref{DM}, and it is propagated by the same TD approach as the deterministic orbitals \cite{White2020, Sharma2023, White2022}.

For our instantaneous electric field pulse along $x$, the vector potential is a step function, ${\bf A}(t) \equiv -\thickhat{e}_{x} E_0\Theta(t)$. Recall that ${\bf E}( t) =- \frac{\partial}{\partial t} {\bf A}(t) $. The electron momentum is thus instantaneously increased and relaxes due to disorder. From the Mermin-DFT approach, the initial current density before the pulse is zero, ${J}(t < 0)=0$.

The force is calculated using the energy-conserving Ehrenfest expression:
\begin{align}
F^{a,y}(f) &=  \int d{\bf r} \left[ \left\{\frac{\partial}{\partial R_{a,y}}  V_{ei, \text{LPP}} ({\bf r}, {\bf R})\right\} \rho({\bf r},t) \right] 
\\\nonumber
&\qquad +\sum_{b,{\bf k}} \int \int d{\bf r} d{\bf r}'  f({\varepsilon^{\bf k}_b, \mu, T}) u^{\bf k *}_b ({\bf r}',t) \left\{\frac{\partial}{\partial R_{a,y}}  V_{ei, \text{NLPP}} ({\bf r},{\bf r}', {\bf R})\right\}   u^{\bf k}_b ({\bf r},t) ~. 
\end{align}

The total electron current can be decomposed into the diamagnetic part, a contribution coming directly from electron density $\rho(r)$ and vector potential, and the paramagnetic part, coming from the time-dependent density matrix, ${\hat \rho}(t)$,
\begin{align}
    {\bf J}(t) &= \int dr {\bf A}(r, t) \rho(r) +  \text{Tr}\left[\left\{\bf \hat p\right\}  {\hat \rho}(t)\right]  ~,
\end{align}
where ${\bf p}=(-i {\bf \nabla} + {\bf k})$ is the canonical electron momentum. Non-local Hamiltonian terms, such as arising from pseudopotentials, add additional terms to the ``full" momentum, as defined via ${\bf p}'= i[{\thickhat H}^{\bf k}_{KS},{\bf r}]$. This ``full" definition retains consistency of the dipole and the momentum operators, while the first fulfils the exact momentum conservation and thus the generalized Dreyer, Coh, and Stengel (DCS) rule as well as the f-sum rule \cite{Resta2018, Dreyer2022}. We employ the canonical momentum in the letter and note that differences are typically less than $\sim$ 10 per cent.

For simplicity, we only discuss response parallel to the direction of perturbation, dropping vector and tensor notation, and  only concern ourselves with the macroscopic $(q\to 0)$ limit. We consider only isotropic systems in the letter and thus, do not investigate the off-diagonal elements of the NBEC or conductivity tensors.

\section{\label{sec:AppendixA2} Current Response to macroscopic electric field pulse and the Drude Weight}
    
The conductivity can be separated into two parts, the regular and Drude contributions, \cite{Resta2018,Bellomia2020}
\begin{align}
    {\bf \sigma} (\omega) &= \frac{D}{\pi} \left\{ \frac{1}{2}\delta(\omega)  + \frac{i}{\omega} \right\} + {\bf \sigma}_{\text{reg}} (\omega) ~.
\end{align}
The Drude part, $\propto D/{\pi}$, and regular conductivity, ${\bf \sigma}_{reg}$, are associated with intraband and interband transitions, respectively. Finite $D$ implies the presence of free charge carriers, \textit{i.e.}, a finite long-time current density   ${\bf J}(t\to \infty)\ne 0$ in response to an instantaneous macroscopic electric field pulse at $t=0$. The electric field imparts momentum to all the electrons equally, but some electrons have high inertia and thus there is a finite current at long times, \textit{i.e.}, timescales of the order of ion motion or phonon frequencies.  Finite simulation time and system size necessitate a dampening of the current before the Fourier transformation, leading to a broadening of the conductivity peaks and a finite regular conductivity at $\omega=0$. By causality, the imaginary part of the conductivity, $\Im [{\bf \sigma} (\omega)]$, is an odd function, so the real part of the  $\Re[i\omega \, {\bf \sigma} (\omega)]\vert_{\omega=0}=0$, therefore from
\begin{align}
    \ i\omega \, {\hat \sigma} (\omega) &\equiv -\frac{{\hat D}}{\pi} + i\omega \, {\hat \sigma}_{reg} (\omega) ~, \quad \text{and} 
    \\
     i\omega \, {\hat \sigma}_{reg} (\omega) &=  \frac{{\bf J}(0^+)}{{\bf E}_0} + \frac{1}{{\bf E}_0}\mathcal{F}\left[\Theta(t)e^{-\frac{1}{4}\gamma^2 t^2}\frac{\partial}{\partial t} {\bf J} (t,{\bf q}=0) \right](\omega) ~,
    \end{align}
we can extract the Drude weight, ${\hat D}/\pi=\Re[i\omega \, {\hat \sigma}_{reg} (\omega)]\vert_{\omega=0}$, directly from the current in the presence of a broadened ($\gamma >0$) regular conductivity. To have the correct $\omega\to \infty$ behavior in $i\omega \, {\hat \sigma} (\omega)$, the dampening term is included after taking the time-derivative.  Since electrons cannot respond instantaneously to the pulse, the current at a time instant infinitesimally after the pulse is proportional to the electric field strength and the electron density, ${\bf J} (0+)  = n_e {\bf E}_0$, which implies, $\int \frac{d\omega }{2\pi} \sigma(\omega)=\frac{n_e}{2}$. \cite{Resta2018}  This is the well-known f-sum rule. More simply put, the Drude weight is proportional to the initial current density plus the (negative) change in the current density over a long time, \textit{i.e.}, it is the \textit{un}relaxed part of the current density.

\section{\label{sec:AppendixB} Validation of approach against systems simulated in ``Dynamical Born effective charges" by C.-Yu Wang, S. Sharma, E. K. U. Gross, and J. K. Dewhurst \cite{Wang2022}} 

\begin{figure*}[h]
    \includegraphics[width=7.in]{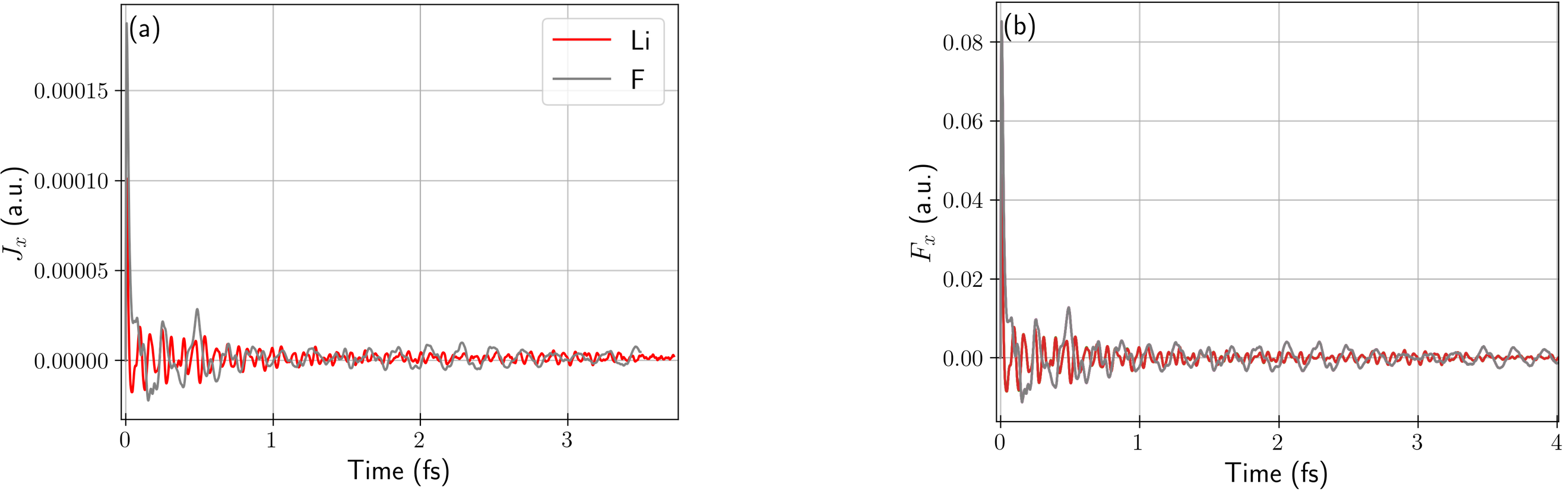}
    \caption{LiF fcc crystal structure with an $8-$atom unit cell and a $[20 \times 20 \times 20]$ Monkhorst-pack $k-$grid: ($a$) Current density signal decay with time for an initial perturbation given by an atomic displacement $\lambda(t=0^-): \delta r = 0.01$ a.u., in the $x-$direction. ($b$) Forces exerted on the atoms by the electronic density in response to an electric field pulse uniform in space in the long-wavelength limit $\lambda (t=0^-): E_x = 0.01$ a.u.}
    \label{fig:signaldecay}
 \end{figure*}

\begin{figure}[h]
    \centering
    \includegraphics[width=7.0 in]{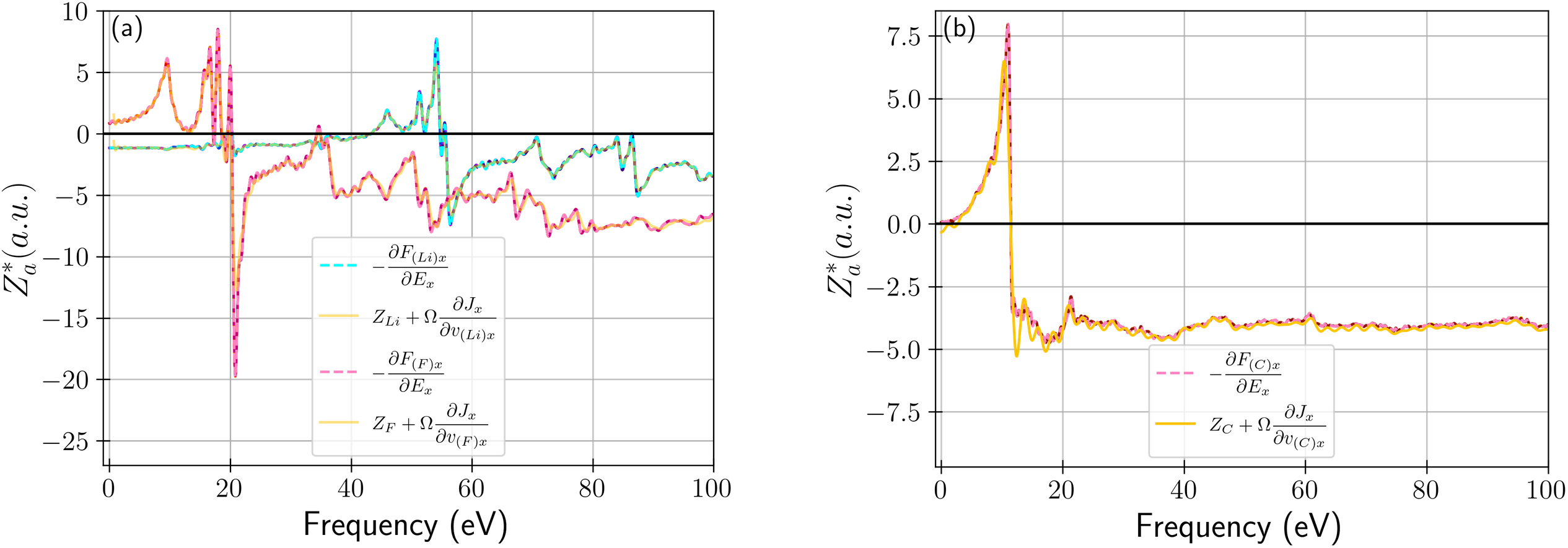}
    \caption{Nonadiabatic Born Effective Charges $Z^*_a$: ($a$) $8-$ atom unit cell of LiF $fcc$ crystal, and ($b$) $8-$ atom unit cell of C diamond, computed using two equivalent definitions of perturbation-response measures.}
    \label{fig:dbec_full}
\end{figure}

A real-time TDDFT-based dynamics, assuming fixed nuclear positions, of the system after an initial perturbation, $\lambda$, reveals the relaxation of the electronic system and the subsequent measure of Born effective charge tensor, $Z^*_{ij}$.
A Gaussian smearing $\gamma$ is used for Fourier transforming the signal to the frequency domain, \textit{vide infra} for values.
In Ref. \cite{Wang2022} the authors use an instantaneous atomic displacement as the initial perturbation and calculated the time-dependent current response. This is based on the relationship:
\begin{align}
    {\hat Z}^{*}_a &= Z_a {\hat I} + \dfrac{\partial \mathbf{P}}{\partial \mathbf{R}^a} \equiv  Z_a {\hat I} + \dfrac{\partial \mathbf{J}}{\partial \mathbf{v}^a} ~, \quad \text {with} \\\nonumber
        \lambda(t) &\equiv \delta \mathbf{R}^a \Theta(t-t_0) = \int_{-\infty}^t dt' \, \mathbf{v}^a (t'), \text{ where } \mathbf{v}^a (t') \equiv \delta \mathbf{R}^a\delta(t'-t_0)  ~,
\end{align}
where $\mathbf{v}^a$ is the velocity of the ion, $a$. $\Theta$ is the Heaviside function. We propose the calculations based on the instantaneous macroscopic electric field pulse:  \begin{align}
    {\hat Z}^{*}_a &= -\dfrac{\partial \mathbf{F}_{a}}{\partial \mathbf{E}} , \quad \text {with} \\\nonumber
    \lambda(t) &\equiv \delta \mathbf{A} \Theta(t-t_0) = \int_{-\infty}^t dt' \, \mathbf{E} (t'), \text{ where } \mathbf{E} (t') \equiv \delta \mathbf{A}\delta(t'-t_0)  ~.
\end{align}
This has the same advantage for calculation of nonadiabatic BEC as it does for the static BEC, the time-dependent propagation (or self-consistent DFPT) can be done for one perturbation, while calculating the observable many times. We validate our approach by comparing with certain calculations carried out in Ref. \cite{Wang2022}, namely ambient conditions lithium fluoride, and diamond, along with aluminium already included in the main text. This also allows us to compare the effect of using semi-core and frozen core pseudopotentials against the all-electron calculations in Ref. \cite{Wang2022}. We calculate similar spectra compared to Ref. \cite{Wang2022}, with large deviations occurring only at large frequencies, where the effect of the pseudopotentials becomes prominent. We see excellent agreement between the two approaches for calculating ${\hat Z}^{*}_a(\omega)$, in both time (Fig. \ref{fig:signaldecay}) and frequency domains (Fig. \ref{fig:dbec_full}).

\section{\label{sec:AppendixD} {Modified Drude-Lorentz model for fitting Group Conductivity and Nonadiabatic Born Effective Charge}}

\begin{align}
\label{eq:drude-plus}
\nonumber
& {\bf \sigma}_{MDL} ([Z_{val}, \Delta Z_{CT},n], \{Z_{eff},{ \tau_0}, \omega_1, {\tau_1},\} , \omega) = \dfrac{\{Z_{eff}+\Delta Z_{CT}\}\, n \, { \tau_0}}{{1 - i \omega{\tau_0}}}  + \frac{\{Z_{val}-Z_{eff}\}n\omega\tau_1}{\omega + i(\omega_1^2-\omega^2)\tau_1} ~,
\\\nonumber
& {{\bf \sigma}}^G_{MDL} (\omega) \equiv \Re \{ {\bf \sigma}_{MDL} ([Z^G_{val}, \Delta Z^G_{CT},n^G], \{Z_{\text{eff}}^G,\tau^G_0,\omega_1^G, \tau^G_1\}, \omega)\} ~,
 \\\nonumber
\\
&  {\bar{Z}}^{*}_{G,MDL} (\omega) \equiv -\Im\{ \omega \times {\bf \sigma}_{MDL} ([Z^G_{val}, \Delta Z^G_{CT},n^G], \{Z_{\text{eff}}^G,\tau^G_0,\omega_1^G, \tau^G_1\}, \omega)\} + \Delta Z^G_{CT} ~.
\end{align}
\new{Here $\tau_0$ and $\tau_1$ are the scattering times for the group's free and bound electrons; $\omega_1$ is the frequency of the bound electrons modeled as a Lorentz oscillator; and $Z_{eff}$ is the free electron number (the average ionization) excluding the number of charge-transferred electrons $\Delta Z_{CT}$.
This maintains the convention that $Z_{val}=Z_{eff}+Z_{bound}$, where $Z_{val}$ is the number of valence electrons that are tightly bound to the nuclei. Parameters $[Z_{val}, \Delta Z_{CT},n]$ are constrained, with $\Delta Z^G_{CT}\equiv {\bar Z}_G^* (0)$, and $n^G$ is the number density for the group atoms. Parameters $\{Z_{eff},{ \tau_0}, \omega_1, {\tau_1},\}$ are the model fit parameters, such that no parameter is negative and $Z_{eff}$ is upper-bound by $Z_{val}$.
When the NBECs are partitioned into groups, the individual ${\bar Z}_G^* (\omega)$ can be finite at zero frequency, even when $D= 0$, due to charge transfer between the partitions. This was shown previously in insulating lithium hydride and boronitride.\cite{Wang2022} In disordered conducting systems the transferred electrons, $\Delta Z^G_{CT}$, may contribute to the conductivity as ``nearly-free" electrons ascribed to a particular group, raising the effective charge. But this charge transfer does not affect the high frequency limit of the NBEC, $Z_a^* (\infty)=Z_{a, val}$, which only depends on $Z_{val}$. Our modified Drude-Lorentz plus charge transfer model, Eq. \eqref{eq:drude-plus} accounts for this.}

\begin{table*}[h]
    \centering
    \caption{\label{tab:tab2} Table of fit parameters for the modified Drude-Lorentz model used to extract $Z^G_{eff}$.}
    \begin{tabular}{ccccccccc} \hline\hline
         Group & $\rho$ [g/cm$^3$] & $k_BT$ [eV]& $Z_{eff}$ & $\tau_0$ (fs) &  $Z_{val}-Z_{eff}$ & $\omega_1$ (eV) & $\tau_1 (fs)$ \\\hline
 \multicolumn{9}{c}{}\\
 Al               & 2.7& 1.0 & 3.0 & 0.48& 0.0 & -- & -- & \\
 C                & 0.5 & 1.0 & 0.55 & 0.08 & 3.45 & 11.6 & 0.05\\
 C                & 3.52& 1.0 & 3.1 & 0.05 & 0.9 & 16.5 & 0.04\\
 All atoms C-Be-H & 1.37& 5.0 & 1.40 & 0.072 & 0.93 & 10.4 & 0.04  \\
 C in C-Be-H      & -- & --& 2.36 & 0.067 & 1.64 & 11.1 & 0.03 \\
 Be in C-Be-H     & -- & --& 1.14 & 0.14  & 0.86 & 8.8 & 0.05 \\
 H in C-Be-H      & -- & --& 0.57 & 0.085 & 0.43 & 9.3 & 0.05 \\

    \end{tabular}
\end{table*}
\new{
For the carbon results the valence includes the four $2s,2p$ electrons which is also the number of electrons treated explicitly in the TD-DFT calculations. For the beryllium results the valence includes only the $2s$ electrons while the $1s$ electrons are also explicitly treated in the TD-DFT calculation. The response of the $1s$ electrons only occurs at $\omega> 100$ eV, and we can safely assume the $1s$ electrons are not ionized for frequencies below this. The fit of the group conductivity for the C-Be-H system ranges from $0-75$ eV, to avoid the Be $1s$ excitations.}

\new{
The fit of the total force-derived conductivity $\sigma^{FD}(\omega)=\big\{\sigma^C(\omega) + \sigma^{Be}(\omega) + \sigma^H(\omega)  \big\}$ can be compared to the individual element group fits. If we had perfect ``nearly-free" Drude contributions only then we would have $Z^{Avg}_{eff}= \frac{1}{3} \big\{ Z^{C}_{eff} + Z^{Be}_{eff} +Z^{H}_{eff} \big \}$. This is approximately satisfied in our system ($1.40$ to $1.36$). We would also have $\tau^{Avg}_0 = \big\{ Z^{C}_{eff}\tau^C_0 + Z^{Be}_{eff}\tau^{Be}_0 +Z^{H}_{eff}\tau^H_0 \big \} / \big\{ Z^{C}_{eff} + Z^{Be}_{eff} +Z^{H}_{eff} \big\} $. Our fit results in $\tau^{avg}_1=0.072$ fs compared to $0.088$ fs. Overlap of the Lorentz oscillators with the Drude contribution, as well as limitations of the fitting model to one oscillator, lead to the differences. 
}

\begin{figure}[h]
    \centering
    \includegraphics[width=7.0 in]{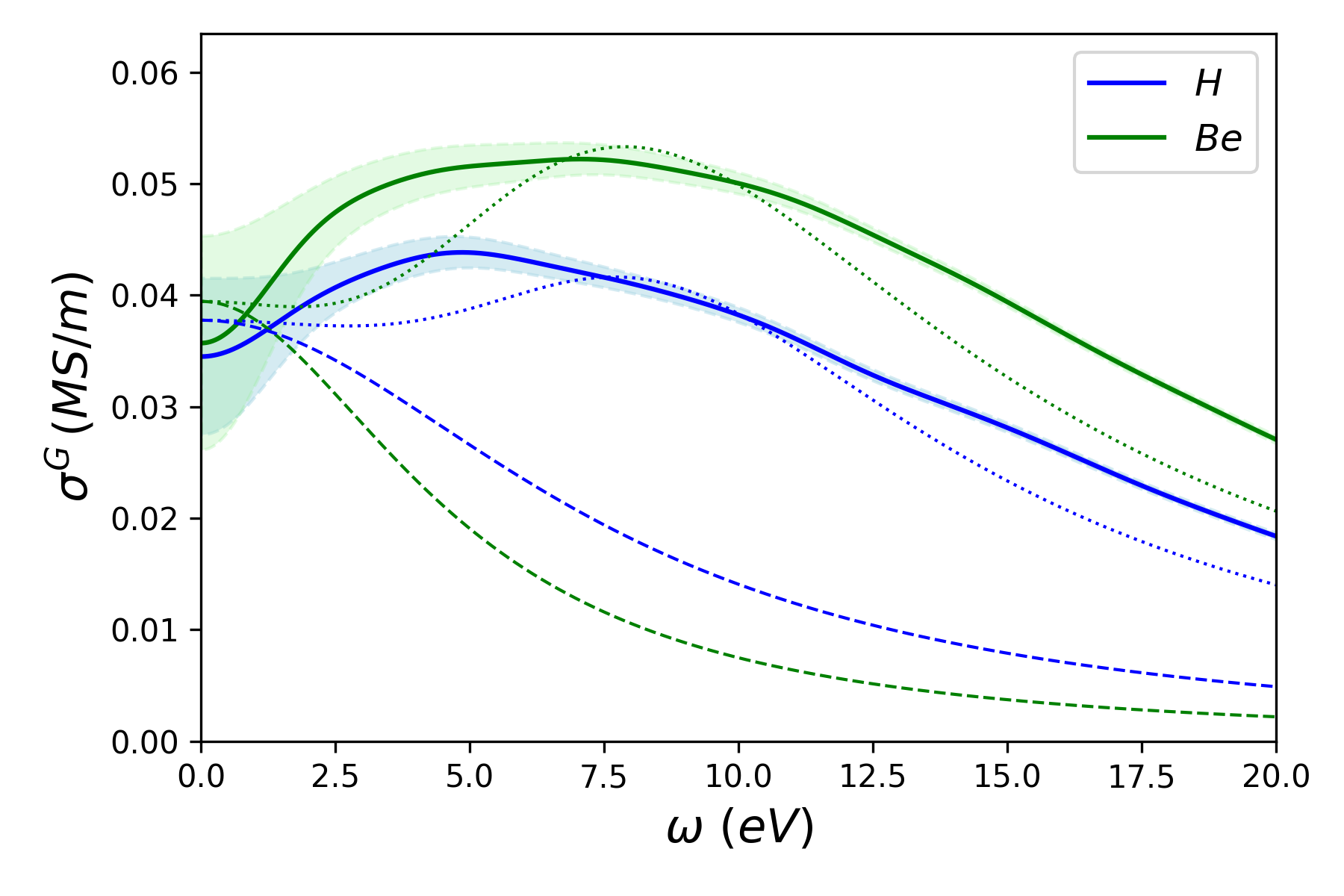}
    \caption{Fitting of C/H/Be mixture (1.37 g/cm$^3$, 5 eV) elemental group conductivities, for H and Be, to the modified Drude-Lorentz model. Solid-Line is the group conductivity, colored band ranges from $\pm$ 1 standard error. Dotted line is modified Drude-Lorentz model fit, dashed line is the Drude only portion (proportional to $Z_{eff}^G$). Low frequency portion is highlighted to show fit and standard error. }
    \label{fig:fits1}
\end{figure}

\begin{figure}[h]
    \centering
    \includegraphics[width=7.0 in]{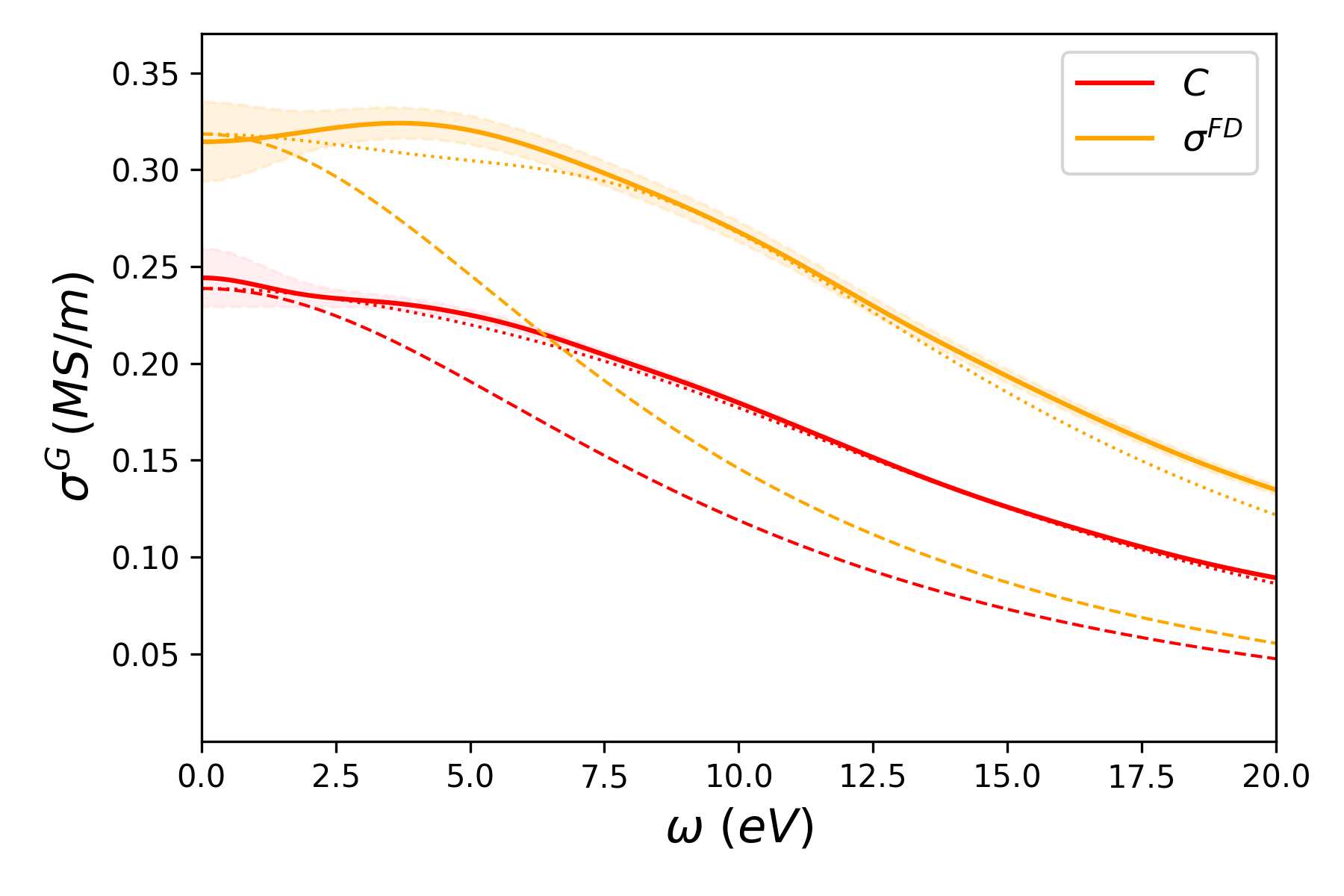}
        \caption{Fitting of C/H/Be mixture (1.37 g/cm$^3$, 5 eV) elemental group conductivities, for C and the total force-derived conductivity ($\sigma^{FD}(\omega)$), to the modified Drude-Lorentz model. Solid line is the group conductivity, colored band ranges from $\pm$ 1 standard error. Dotted line is modified Drude-Lorentz model fit, dashed line is the Drude only portion (proportional to $Z_{eff}^G$). Low frequency portion is highlighted to show fit and standard error.}
    \label{fig:fits2}
\end{figure}
\section{\label{sec:AppendixC} Simulation details}

All simulations in the letter and these supplemental materials are performed via time-dependent Kohn-Sham density functional theory. We apply either fully deterministic or mixed stochastic-deterministic approaches, using the pseudopotential plane-wave DFT code \texttt{SHRED} (Stochastic and Hybrid Representation of Electronic structure by Density functional theory) developed by the authors at Los Alamos National Laboratory \cite{White2020, White2022}. Table \ref{tab:tab1} shows the parameters used to perform the simulations. 
We utilize the Short Iterative Lanczos propagation scheme \cite{Park1986}, with no enforced time reversal symmetry. For section \ref{sec:AppendixB}, we utilize Hartwigsen-Goedecker-Hutter (HGH) pseudopotentials from the \texttt{cp2k} database \cite{Goedecker1996}. For the simulations presented in the letter, we utilize Optimized Norm-Conserving Vanderbilt pseudopotentials \cite{Hamann2013} (ONCV, version 3.2.3) from the PseudoDojo repository \cite{Setten2018}.  \new{For Figure 1 of the letter the FCC results are compared to a single disordered ion configuration taken from molecular dynamics simulation. For Figures 2(3), we average over 5(4) ion configurations sampled from molecular dynamics simulations, performed using the SHRED code.To generate the atomic configurations for WDM systems, we carried out Born-Oppenheimer molecular dynamics (BOMD) simulations in an isokinetic (canonical NVT)\cite{Minary:2003}ensemble using a GGA-PBE exchange-correlation functional; the time-step is based on Wigner-Seitz radius and temperature. After an initial equilibration period, each snapshot is sampled from the production BOMD with sufficient period to ensure uncorrelated snapshots. These snapshots were further used as structures for the TD-DFT real-time dynamics.}
\begin{table*}[h]
    \centering
    \caption{\label{tab:tab1} Simulation parameters for the letter and supplemental materials. Key: $\rho$ - total mass density, $k_{B}T$ - electron temperature, $N_{\psi}$ - number of deterministic Kohn Sham orbitals, $N_\chi$ - number of complementary stochastic vectors, k-grid - Broullion Zone sample via Monkhorst-Pack grids (no $\Gamma$) \cite{Monkhorst1976}, $N_a$ - number of atoms in unit cell, Ecut - maximum planewave energy defining basis and real-space grid, $\gamma$ - the Gaussian dampening coefficient / broadening parameter, $E_0$ - the perturbing electric field strength, $dt$ the electronic time step.    }
    \begin{tabular}{ccccccclccc} \hline\hline
         System & $\rho$ [g/cm$^3$] & $k_BT$ [eV]& $N_\psi$ & $N_{\chi}$& k-grid& $N_{a}$&  Ecut [eV]&$\gamma$ [eV]&  $E_0$[a.u.]& $dt$ [a.u.] \\\hline
 \multicolumn{11}{c}{}\\
 \multicolumn{11}{c}{--Letter--}\\ 
         Al (FCC)& 2.7& 0.025& 14& 0& $16\times16\times16$ & 4&  500&0.136& 0.01& 0.05\\
         Al (FCC)& 2.7& 1.0& 28& 0& $16\times16\times16$ & 4&  500&0.136& 0.01& 0.05\\
         Al (FCC)& 2.7& 3.0& 42& 0& $16\times16\times16$ & 4&  500&0.136& 0.01& 0.05\\ 
 Al (FCC)& 2.7& 5.0& 84& 0& $16\times16\times16$ & 4&  500&0.136& 0.01&0.05\\
 Al (WDM)& 2.7& 1.0& 224& 0& $2\times2\times2$ & 64&  500&0.015& 0.01&0.05\\
 C (WDM)& 0.5& 1.0& 320& 0& $\Gamma-$point& 64&  \cut{987}\new{1000} &0.272& 0.01&0.022\\
 C (WDM)& 3.52& 1.0& 224& 0& $2\times2\times2$ & 64& 1000& 0.272& 0.01&0.021\\
 C-Be-H (WDM)& 1.37& 5.0& 1120& 112& $\Gamma-$point& $128/128/128$ & 1200& \cut{0.272} \new{1.36}& 0.1&0.018\\
 \multicolumn{11}{c}{}\\\hline
 \multicolumn{11}{c}{--Supplemental--}\\
 Li-F (fcc)& $2.53$ & $0.025$ & $24$ & $0$ & $20\times20\times20$ & $4/4$ & $500$ & $0.136$ & $0.01$ & $0.043$\\
 C (Diamond)& $3.53$ & $0.025$ & $16$ & $0$ & $20\times20\times20$ & $4$ & $500$ & $0.136$ & $0.01$ & $0.043$\\
    \end{tabular}
\end{table*}

\end{document}